\documentclass[manuscript,screen]{acmart}

\usepackage{xcolor}
\usepackage{mathtools}
\usepackage{multirow}
\usepackage{subcaption}
\usepackage{algorithm}
\usepackage{algorithmic}

\graphicspath{{images/}}
\DeclareGraphicsExtensions{.pdf}

\DeclarePairedDelimiter{\set}{\{}{\}}
\DeclarePairedDelimiter{\tuple}{(}{)}
\DeclarePairedDelimiter{\abs}{\lvert}{\rvert}
\renewcommand{\implies}{\rightarrow}
\newcommand{\db}{D}
\newcommand{\dbs}{\mathcal{D}}

\newcommand{\mi}[2]{I(#1;#2)}

\begin{document}

\title{Quantitative Auditing of AI Fairness with Differentially Private Synthetic Data}

\author{Chih-Cheng Rex Yuan}
\email{hello@rexyuan.com}
\affiliation{%
  \institution{Institute of Information Science, Academia Sinica}
  \city{Taipei}
  \country{Taiwan}
}

\author{Bow-Yaw Wang}
\email{bywang@iis.sinica.edu.tw}
\affiliation{%
  \institution{Institute of Information Science, Academia Sinica}
  \city{Taipei}
  \country{Taiwan}
}



\begin{abstract}
Fairness auditing of AI systems can identify and quantify biases. However, traditional auditing using real-world data raises security and privacy concerns. It exposes auditors to security risks as they become custodians of sensitive information and targets for cyberattacks. Privacy risks arise even without direct breaches, as data analyses can inadvertently expose confidential information. To address these, we propose a framework that leverages differentially private synthetic data to audit the fairness of AI systems. By applying privacy-preserving mechanisms, it generates synthetic data that mirrors the statistical properties of the original dataset while ensuring privacy. This method balances the goal of rigorous fairness auditing and the need for strong privacy protections. Through experiments on real datasets like Adult, COMPAS, and Diabetes, we compare fairness metrics of synthetic and real data. By analyzing the alignment and discrepancies between these metrics, we assess the capacity of synthetic data to preserve the fairness properties of real data. Our results demonstrate the framework's ability to enable meaningful fairness evaluations while safeguarding sensitive information, proving its applicability across critical and sensitive domains.
\end{abstract}

\ccsdesc[500]{General and reference~Metrics}
\ccsdesc[500]{Security and privacy~Usability in security and privacy}
\ccsdesc[300]{Information systems~Data mining}
\ccsdesc[300]{Computing methodologies~Machine learning}
\ccsdesc[300]{Computing methodologies~Artificial intelligence}
\ccsdesc[300]{Security and privacy~privacy-preserving protocols}
\ccsdesc[300]{Information systems~Information systems applications}
\ccsdesc[300]{Mathematics of computing~Contingency table analysis}

\keywords{AI, fairness, auditing, differential privacy, synthetic data}


\maketitle

\section{Introduction}



Fairness in machine learning has become an increasingly important topic as AI systems are widely used across various sectors, including healthcare and criminal justice. These AI systems can significantly influence outcomes that impact individuals and communities.

However, AI systems can exhibit bias, which may lead to increased unfairness. Biased AI systems have the potential to reinforce historical injustices, erode trust in technology, and exacerbate inequality. Therefore, ensuring fairness in AI systems is crucial to prevent the perpetuation of these biases.

For instance, investigative journalists at ProPublica uncovered bias in the COMPAS system, which is used in the United States criminal justice system to predict recidivism. Their analysis revealed that the COMPAS system disproportionately discriminated against certain demographic groups.

These findings have raised significant concerns about the fairness and ethical implications of using such AI systems in critical decision-making processes. Addressing these biases is essential to ensure that AI systems do not perpetuate inequalities.


One effective approach to achieving fairness in AI systems is to conduct audits. Quantitative fairness auditing involves evaluating these systems using various fairness metrics to identify potential biases and inequalities. Some commonly proposed metrics for assessing AI system fairness include demographic parity, equalized odds, and accuracy equality. These metrics provide a quantitative basis for addressing fairness concerns in AI models.

A well-documented example of fairness auditing is the investigation of the COMPAS system. Journalists from ProPublica revealed that COMPAS exhibited bias against certain demographic groups, particularly African Americans. The system was more likely to falsely classify African American defendants as high risk compared to their Caucasian counterparts. This finding not only underscored the necessity of fairness auditing but also highlighted the potential harm of deploying biased AI systems in critical areas.

This case emphasizes the importance of independent audits in ensuring fairness. Evaluations conducted by third parties, such as those performed by ProPublica, are vital because they operate free from the influence of those who develop or deploy the systems. By offering an objective perspective, third-party audits can identify and address biases that internal evaluations might overlook. Without such oversight, AI systems risk perpetuating systemic inequalities, making independent auditing a crucial aspect of accountability in AI development.



While fairness auditing is essential, it presents significant challenges, particularly when it involves sensitive data. To conduct audits, auditors may rely on access to real datasets containing sensitive information. Auditors often require access to real-world datasets containing sensitive attributes, such as race, gender, or income, to evaluate an AI system's fairness. However, this reliance on sensitive data raises concerns about data security and privacy.

This reliance invites data security attacks on auditors, making them targets of unauthorized access. The use of real datasets for fairness auditing creates a point of vulnerability, making auditors potential targets for data breaches and unauthorized access. If sensitive data falls into the wrong hands, it could lead to severe consequences, including identity theft, discrimination, and reputational damage.

Moreover, analyses of real data run the risk of data inference attacks, where confidential information may be deduced from seemingly harmless or aggregated data. Even if the real data is not directly exposed, the results of analyses can sometimes reveal sensitive information through inference attacks. For example, attackers could combine publicly available audit results with external datasets to infer private details about individuals in the original dataset.


Various technologies have been developed to mitigate privacy risks, such as differential privacy. Differential privacy is a mathematical framework that ensures the inclusion or exclusion of any individual in a dataset has a minimal and quantifiable impact on the overall results of data analysis. This is achieved by introducing carefully calibrated random noise into the computations derived from the data, ensuring that individual records remain indistinguishable while preserving aggregate patterns and trends.

Differential privacy is especially useful in situations that involve sensitive data. For example, a differential privacy mechanism could allow researchers to analyze patterns in patient health data without exposing any single patient's medical history. This approach strikes a balance between data utility and privacy, enabling meaningful analysis while safeguarding individuals against privacy breaches or inference attacks.

The key idea is that no matter how much background knowledge an attacker has about the data, they cannot determine with high confidence whether a specific individual is included in the dataset. This is quantified by some parameters which control the amount of noise introduced; smaller parameters provide stronger privacy but can reduce the accuracy of the analysis, while larger parameters allow for more precise results but with weaker privacy guarantees. The challenge is to carefully calibrate the parameters to ensure that privacy is maintained without overly compromising the utility of the data.


One way of auditing while preserving privacy is to use synthetic data. Synthetic data are fake data that is generated from real data. Synthetic data is artificially generated to replicate the statistical properties of real data while protecting privacy. The principle of this technology is to preserve some statistical properties of the real data while protecting the privacy of the individuals in the real data. Unlike anonymized data, synthetic data breaks the link to the original data, reducing the risk of re-identification.

Among the generation techniques, differentially private synthetic data offers a way to investigate the data properties under differential privacy guarantees. It stands out for it strong differential privacy guarantees when generating synthetic data. This method combines data synthesis with the principles of differential privacy, ensuring that individuals' sensitive information cannot be inferred from the synthetic data. By introducing controlled randomness during generation, it provides strong privacy guarantees while preserving the overall utility of the data.

Differentially private synthetic data is particularly valuable in sensitive domains like healthcare and criminal justice, enabling researchers to analyze patterns or identify trends without exposing personal details. It also supports fairness auditing by allowing the examination of datasets and models without revealing sensitive attributes. This dual focus on privacy and utility makes it a valuable tool.



We propose a novel auditing framework that leverages differentially private synthetic data to evaluate the fairness of AI systems. Our framework lets the auditors assess the fairness of AI systems without exposing sensitive information. By using differentially private synthetic data, auditors can conduct fairness analyses without directly exposing sensitive information.

We present an innovative solution to the issue of sensitive data access: auditors generate synthetic data immediately upon receiving the real dataset, which they retain for auditing purposes while discarding the original data.

This method offers a unique solution to reconcile the need for fairness auditing with the imperative to protect privacy. By preserving key statistical properties of the original data, synthetic data allows auditors to assess fairness metrics accurately while mitigating the risks of data breaches and privacy violations.

This technology enables auditing without exposing sensitive information, offering a privacy-preserving way to assess AI systems. However, synthetic data also introduces new challenges. One major concern is whether it accurately reflects the fairness properties of the real data. Whether or not it accurately preserves the fairness properties of the real data is a critical question.

Since fairness metrics depend on the relationships between sensitive attributes and outcomes, any discrepancies in how synthetic data mimics these relationships could lead to inaccurate conclusions in fairness audits. Ensuring that synthetic data preserves these properties is critical for reliable auditing.

Our work examines the capabilities of existing differentially private synthetic data generation technologies in preserving fairness properties. Through empirical experiments on multiple real datasets, we aim to determine the effectiveness of differentially private synthetic data in fairness auditing.

We perform a comparative analysis of fairness metrics computed on real and synthetic data across multiple real-world datasets, including Adult, COMPAS, and Diabetes. By examining the consistency and differences between these metrics, we evaluate how well synthetic data captures the fairness characteristics of real data.

Through our experiments, we evaluate the extent to which synthetic data can approximate fairness metrics computed on original datasets. Our findings contribute to the growing body of research on ethical AI and offer insights for practitioners seeking to ensure fairness in AI systems.


The rest of this paper is organized as follows. Section 2 provides the background of related work. Section 3 provides the preliminary knowledge. Section 4 outlines the motivation behind our proposed framework. Section 5 describes our methodology for generating differentially private synthetic data and evaluating fairness metrics. Section 6 presents our experimental setup and results. Section 7 discusses the implications of our findings. Section 8 concludes the paper and outlines future work.

\section{Related Work}


Fairness in machine learning has become a crucial area of research\cite{barocas2023fairness}. Various fairness measures have been proposed to quantify\cite{yeh2024analyzing} the fairness of AI models\cite{pessach2022review,corbett2017algorithmic,vzliobaite2017measuring,hardt2016equality,corbett2017algorithmic,berk2021fairness,chouldechova2017fair,kleinberg2016inherent}. We consider these fairness measures in this work.

Auditing of AI models is an important step in ensuring that these models are fair\cite{ferrara2023fairness}. Different tools have been developed to check the fairness of AI models\cite{saleiro2018aequitas,bellamy2019ai,bird2020fairlearn}. We use a tool produced from \cite{yuan2024ensuring}.


Differential privacy has risen to become the standard of privacy protection in many data analyses\cite{jiang2021applications}. It offers a formal framework for quantifying privacy guarantees when releasing information derived from sensitive data\cite{dwork2006calibrating,dwork2014algorithmic}. We use differential privacy in this work to protect the privacy of individuals in the data in our auditing framework.

Different flavors of differential privacy have been proposed over the years\cite{desfontaines2019sok}, such as Gaussian differential privacy\cite{dong2022gaussian}, Pufferfish differential privacy\cite{kifer2012rigorous}, Bayesian differential privacy\cite{triastcyn2020bayesian}, and R\'enyi differential privacy\cite{mironov2017renyi}. We consider R\'enyi differential privacy specifically in this work.


The generation of synthetic data\cite{raghunathan2021synthetic,lu2023machine} under differential privacy constraints allows for data sharing without compromising privacy\cite{tao2021benchmarking}. There have been several techniques developed for differentially private synthetic data generation\cite{rosenblatt2020differentially,fan2020relational,bowen2019comparative,bowen2020comparative,arnold2020really,xu2019modeling} in recent years. We use the technique that won the 2018 NIST differential privacy Synthetic Data Challenge competition\cite{mckenna2021winning}.

There have also been some works that aim to alter the properties of synthetic data, such as introducing bias to them\cite{jiang2024synthetic,baumann2023bias}. In this work, we aim to preserve the properties of the original data. Thus, we do not consider these works.

While helpful, there are some technical limitations\cite{stadler2022synthetic,cheng2021can,ganev2022robin,wyllie2024fairness} and ethical concerns\cite{whitney2024real} with the use of synthetic data. Synthetic data may not always preserve properties of real data. This may cause inaccuracies in downstream tasks such as fairness evaluation. Our work examines these limitations.

\section{Preliminaries}
\label{sec:prelim}

Let $\prod_i x_i$ denote the Cartesian product of $x_i$s.

An \emph{attribute} is a symbol $A$.
The set of all attributes is $\mathcal{A}$.
An \emph{attribute value} is a symbol $a$.
The set of all attribute values is $\Omega$.
An \emph{attribute value space} is a function $\sigma : \mathcal{A} \implies 2^{\Omega}$
specifying the set of valid values that attributes can take on.
A \emph{row} with respect to $\sigma$
is a function $r : \mathcal{A} \implies \Omega$
where $r(A) \in \sigma(A)$.
The set of all rows is $\mathcal{R}$.
A \emph{database} is a tuple $\db =
\tuple{
    \mathcal{A}_{\db},
    \Omega_{\db},
    \sigma_{\db},
    \mathcal{R}_{\db}
}$
where
$\mathcal{A}_{\db} \subseteq \mathcal{A}$
is the set of attributes in $\db$,
$\Omega_{\db} \subseteq \Omega$
is the set of attribute values in $\db$,
$\sigma_{\db} : \mathcal{A}_{\db} \implies 2^{\Omega_{\db}}$
is the function specifying the valid values that each attribute can take on in $\db$,
and $\mathcal{R}_{\db} \subseteq \mathcal{R}$
is the set of rows with respect to $\sigma_{\db}$ in $\db$
with $r_\db : \mathcal{A}_\db \implies \Omega_\db$
for all $r_\db \in \mathcal{R}_\db$
and $r_\db(A) \in \sigma_\db(A)$
for all $A \in \mathcal{A}_\db$.
The set of all databases is $\dbs$.

Let $X \in \mathcal{X},Y \in \mathcal{Y}$ be discrete random variables. Their \emph{mutual information} $\mi{X}{Y}$ is
\[
\mi{X}{Y} = \sum_{x \in \mathcal{X}} \sum_{y \in \mathcal{Y}} P_{(X,Y)}(x,y) \text{log} \frac{P_{(X,Y)}(x,y)}{P_X(x)P_Y(y)}
\]
where $P_{(X,Y)}$ is the joint probability distribution of $X,Y$ and $P_X,P_Y$ are the marginal probability distributions of $X,Y$ respectively.
It quantifies the level of dependency between $X$ and $Y$.

Let $\set{X_i}$ be a set of discrete random variables indexed by a graph $G = (V,E)$, where $V$ represents the random variables $X_i$s and $E$ represents dependencies between these random variables. A \emph{Markov random field} is a probability distribution over $X_i$s, such that each random variable $X_i$, given its neighborhood in $G$, is conditionally independent of all other variables. Since edges represent dependencies, cliques in a Markov random field represent groups of variables that are all mutually dependent. As a machine learning model, there has been much development in the estimation and inference of Markov random fields\cite{koller2009probabilistic,murphy2023probabilistic}.

\subsection{Fairness Measures}

For fairness measures\cite{yuan2024ensuring,pessach2022review}, let $Y$ denote the ground truth of an outcome, let $\hat{Y}$ denote the predicated result of an outcome, let $S$ denote the protected attribute, and let $\epsilon$ denote some threshold. $Y, \hat{Y}, S$ are binary.

Fairness measures can be broadly categorized into independence, separation, and sufficiency, which are defined by conditional independence in Table~\ref{tab:categories}. $X \bot Y | Z$ denotes the conditional independence between $X$ and $Y$ conditioning on $Z$.

\begin{table}[h]
\caption{Fairness categories.}
\label{tab:categories}
\begin{tabular}{cc}
\toprule
\textbf{Category} & \textbf{Definition} \\
\midrule
Independence & $S \bot \hat{Y}$ \\
Separation & $S \bot \hat{Y} | Y$ \\
Sufficiency & $S \bot Y | \hat{Y}$ \\
\bottomrule
\end{tabular}
\end{table}

These categories can be expanded into forms of probability. For example, the definition of separation is expanded to
\begin{align*}
P[\hat{Y} = 1 | S = 1, Y = 1] & = P[\hat{Y} = 1 | S \neq 1, Y = 1] \\
P[\hat{Y} = 1 | S = 1, Y = 0] & = P[\hat{Y} = 1 | S \neq 1, Y = 0]
\end{align*}
The definition can be relaxed. Its relaxation, for some parameter $\epsilon$, is
\begin{align*}
\abs{P[\hat{Y} = 1 | S = 1, Y = 1] - P[\hat{Y} = 1 | S \neq 1, Y = 1]} & \leq \epsilon \\
\abs{P[\hat{Y} = 1 | S = 1, Y = 0] - P[\hat{Y} = 1 | S \neq 1, Y = 0]} & \leq \epsilon
\end{align*}
which is also the definition of a fairness measure called equalized odds.

We consider in this work various fairness measures listed in Table~\ref{tab:measures}.

\begin{table*}[h]
\caption{Fairness measures.}
\label{tab:measures}
\begin{tabular}{lll}
\toprule
\textbf{Category} & \textbf{Fairness Measure} & \textbf{Definition} \\
\midrule
\multirow{1}{*}{Independence}
& Demographic Parity & $\abs{P[\hat{Y} = 1 | S = 1] - P[\hat{Y} = 1 | S \neq 1]} \leq \epsilon$ \\
\multirow{2}{*}{Separation}
& \multirow{1}{*}{Equalized Odds (False Positive)} & $\abs{P[\hat{Y} = 1 | S = 1, Y = 0] - P[\hat{Y} = 1 | S \neq 1, Y = 0]} \leq \epsilon$ \\
& \multirow{1}{*}{Equalized Odds (True Positive)} & $\abs{P[\hat{Y} = 1 | S = 1, Y = 1] - P[\hat{Y} = 1 | S \neq 1, Y = 1]} \leq \epsilon$ \\
\multirow{2}{*}{Sufficiency}
& \multirow{1}{*}{Conditional Use Accuracy Equality (True Positive)} & $\abs{P[Y = 1 | S = 1, \hat{Y} = 1] - P[Y = 1 | S \neq 1, \hat{Y} = 1]} \leq \epsilon$ \\
& \multirow{1}{*}{Conditional Use Accuracy Equality (True Negative)} & $\abs{P[Y = 0 | S = 1, \hat{Y} = 0] - P[Y = 0 | S \neq 1, \hat{Y} = 0]} \leq \epsilon$ \\
\multirow{1}{*}{N/A}
& Overall Accuracy Equality & $\abs{P[Y = \hat{Y} | S = 1] - P[Y = \hat{Y} | S \neq 1]} \leq \epsilon$ \\
\bottomrule
\end{tabular}
\end{table*}

\subsection{R\'enyi differential privacy}


A \emph{randomized mechanism} is a randomized algorithm $M : \dbs \implies \mathbb{R}^p$ that takes a database and, after introducing noise, outputs some results.

Let $\db_1,\db_2$ be two databases. They are \emph{neighbors}, denoted $\db_1 \sim \db_2$, if $\abs{\mathcal{R}_{\db_1} \Delta \mathcal{R}_{\db_2}} = 1$ where $\Delta$ denotes symmetric difference; that is, one database contains an extra row than the other. In other words, they differ in exactly one row.

\begin{definition}[Gaussian Mechanism\cite{dwork2014algorithmic}]\label{def:gm}
Let $f : \dbs \implies \mathbb{R}^p$ be a function. The Gaussian Mechanism $M$ adds independent and identically distributed Gaussian noise with mean $0$ and standard deviation $\sigma$ to each component of the $p$-dimensional vector output of $f(\db)$
\[
M(\db) = f(\db) + \mathcal{N}(0_p, \sigma^2 \mathbb{I}_p)
\]
where $\mathcal{N}$ is a multivariate normal distribution with mean vector $0_p$ and covariance matrix $\sigma^2 \mathbb{I}_p$ where $\mathbb{I}_p$ is the identity matrix.
\end{definition}

\begin{definition}[R\'enyi Differential Privacy (RDP)]\label{def:rdp}
Let $P_{\mathbf{X}}$ denote the probability distribution associated with the random vector $\mathbf{X}$. A randomized mechanism $M$ satisfies $(\alpha,\gamma)$-RDP for $\alpha \geq 1$ and $\gamma \geq 1$ if, for all databases $\db_1 \sim \db_2$, we have
\[
D_\alpha(P_{M(\db_1)} \Vert P_{M(\db_2)}) \leq \gamma
\]
where $D_\alpha(P_1 \Vert P_2)$ is the R\'enyi divergence\cite{van2014renyi,van2010renyi,li2016renyi} of order $\alpha$ between probability distributions $P_1,P_2$ over $x$
\[
D_\alpha(P_1 \Vert P_2) := \frac{1}{\alpha - 1} \text{log} \int {P_1(x)}^\alpha {P_2(x)}^{1-\alpha} \text{d}x
\]
\end{definition}

\begin{theorem}[RDP of the Gaussian Mechanism\cite{feldman2018privacy,mironov2017renyi}]
The Gaussian Mechanism satisfies $(\alpha, \alpha \frac{\Delta^2_f}{2 \sigma^2})$-RDP, where $\Delta_f$ denotes the sensitivity\cite{dwork2014algorithmic} of $f$, which is defined as the maximum $L^2$-norm difference in the output of $f$
\[
\Delta_f := \text{max}_{\db_1 \sim \db_2} \Vert f(\db_1) - f(\db_2) \Vert_2
\]
\end{theorem}

\subsection{Differentially Private Synthetic Data}

Let $C \subseteq \mathcal{A}$ be a subset of attributes. Let $\Omega_C = \prod_{i \in C} \Omega_i$. Let $x$ be a row and $x_C$ denote the restriction of $x$ to $C$. A \emph{marginal}\cite{barak2007privacy,mckenna2021winning} of $C$ on database $\db$ is a function $\mu_{\db} : \Omega_C \implies \mathbb{N}_0$ such that $\mu_{\db}(t) = \Sigma_{x \in \mathcal{\db}} \delta_{t,{x_C}}$ where $\delta$ is the Kronecker function; that is, it is a lookup table of the counts of each possible combination of attribute values. We call marginals of $|C| = n$ attributes $n$-way marginals.

The task of differentially private synthetic data\cite{Ullman2022,McKenna2022,NIST_Differentially_Private_Synthetic_Data,YouTube_Differentially_Private_Synthetic_Data} is, given a database $\db$, adding some noise to marginals of $\db$ such that it satisfies some differential privacy guarantees and outputting another database $\db'$, such that the $L^1$-norm errors between marginals of $\db$ and $\db'$ is small; that is, their marginals $\mu_{\db},\mu_{\db'}$ are similar.

For example, suppose we have a database with attributes sex and race. The 2-way marginals of the original database and the synthetic database are shown in Table~\ref{tab:marginal-example}. The marginals of the synthetic data is supposed to be similar to that of the original database.

\begin{table}[h]
\caption{Example marginals.}
\label{tab:marginal-example}
\centering
\begin{subtable}[t]{0.45\linewidth}
\centering
\caption{Marginal of original data.}
\begin{tabular}{lr}
\toprule
Attributes & Count \\
\midrule
Male,White & 24 \\
Female,White & 33 \\
Male,Black & 13 \\
Female,Black & 47 \\
\bottomrule
\end{tabular}
\end{subtable}
\begin{subtable}[t]{0.45\linewidth}
\centering
\caption{Marginal of synthetic data.}
\begin{tabular}{lr}
\toprule
Attributes & Count \\
\midrule
Male,White & 22 \\
Female,White & 35 \\
Male,Black & 10 \\
Female,Black & 46 \\
\bottomrule
\end{tabular}
\end{subtable}
\end{table}

\section{Motivation}

AI systems are widely used in various aspects in our society. However, these systems may be biased against certain demographic groups. For example, COMPAS, the AI system used in American criminal justice system to predict recidivism, has been found to be biased by ProPublica, a third party media.

To ensure the fairness of AIs, third-party audits such as this is crucial. Auditing can reveal instances of unfairness in AI systems, allowing for corrective actions to be taken. We hence advocate for a framework that centers around third-party audits.

Our auditing framework is tripartite. It consists of three parties: the data provider, the model maker, and the third-party auditor. The COMPAS investigation could fit in with this framework.

The data provider is responsible for supplying the raw datasets which should originate from trustworthy sources, such as government agencies like a census bureau. In the COMPAS example, the data provider is the government agencies that provide criminal justice data, such as the Broward County Sheriff's Office in Florida.

The model maker develops AI models. These are AI companies or research labs specialized in training and optimizing AI models. The model maker in the COMPAS example is Northpointe, Inc., the company that developed the COMPAS system.

The third-party auditor acts as an evaluator, using our tool to audit the AI models for fairness issues by combining both the datasets and the models. These may be investigative journalists or regulatory bodies. ProPublica is the third-party auditor in the COMPAS example, using quantitative methods to evaluate the fairness of the COMPAS system using Florida's data.

\begin{figure}[h]
  \centering
  \includegraphics[width=\linewidth]{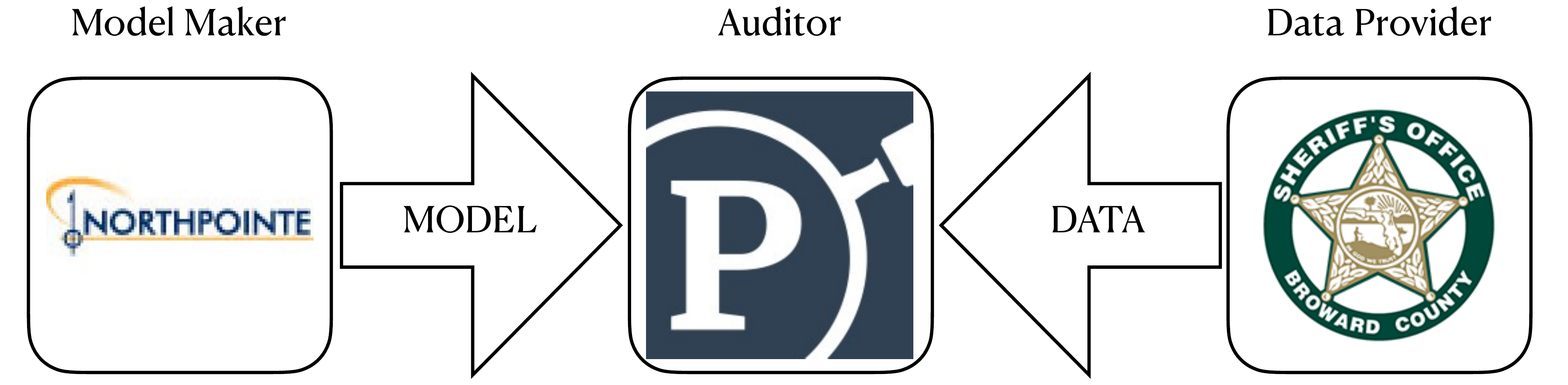}
  \caption{Roles of each party in the COMPAS example.}
  \label{fig:compas}
  \Description{Roles of each party in the COMPAS example.}
\end{figure}

In the framework of \cite{yuan2024ensuring}, after obtaining real data from the data provider, the 3rd party auditor holds onto the real data for performing fairness audits, and it supposedly retains it indefinitely for the possibility of any future audits. However, this practice raises both security and privacy concerns.

For security, it creates a point of vulnerability to unauthorized access, and the auditor is now a target of data security attacks. The auditor may not have the necessary resources to defend against these threats. A breach at the auditor's end could result in compromises of individuals' sensitive information. In the COMPAS example, had the data not been public already, ProPublica may be targeted by hackers to access the sensitive criminal justice data.

As for privacy, on the other hand, it introduces risks of information leakage. Releasing analyses of real data may inadvertently reveal sensitive information through data inference attacks. Attackers can exploit data patterns in outputs or combine outputs with external data to infer sensitive information. In the COMPAS example, had the data not been public already, their analyses may inadvertently reveal the identities of individuals in the criminal justice system.

Thus, we introduce a new framework where the auditor generates synthetic data based on real data upon retrieval of the real data, and then holds onto the synthetic data and discards the real data, preventing all further security and privacy violations. The third-party still retains the ability to audit all incoming future models as needed.

\section{Methodology}
\label{sec:method}

We employed the tools of the winner of the 2018 NIST differential privacy Synthetic Data Challenge competition\cite{NIST2018} by \cite{nist_ryan_2018,McKenna_private-PGM_2021,mckenna2021winning,mckenna2019graphical} and the fairness checker tool from \cite{yuan2024ensuring}.

We shall illustrate the workflow of our methodology with an example of the COMPAS dataset, which contains attributes such as sex and race with the prediction target attribute being recidivism.

\subsection{Data Synthesis}

The synthesis framework is three-fold, namely, select-measure-generate\cite{McKenna2022,YouTube_Differentially_Private_Synthetic_Data}. We first select the important marginals to preserve, measure them by adding differential privacy noise, and then generate synthetic data.

Underneath the hood, the tool employs a Markov random field. The select step corresponds to marking cliques in a Markov random field, and the generate step corresponds to sampling from the fitted Markov random field.

By default, all $1$-way marginals are selected to preserve the quantity of each attribute element. We can further preserve correlations by adding $n$-way marginals. For example, if we want to preserve the relationship between sex and race, we may add the clique (sex,race).

In a perfect world where all correlation information is to be preserved, we may wish to make a complete graph. However, this is intractable as the complexity of the problem would skyrocket. Furthermore, algorithms for Markov random field favor graphs with specific shapes, such as trees.

To circumvent this complexity explosion, instead, \cite{mckenna2021winning} devised a technique where the mutual information of all the database attribute pairs is calculated, and then a maximum spanning tree is identified with edge weights being the mutual information to obtain a skeleton tree-shaped Markov random field. The marginals are then selected based on the edges of this tree.

For the competition, \cite{mckenna2021winning} further manually added certain cliques based on their investigation of the competition dataset. For example, they would add some edges based on some sophisticated heuristics tailored to that particular dataset. In addition, they manually added some particular cliques like (sex,city,income) for the prediction target of income. We do not consider these additional optimizations for the generality of our approach. Meanwhile, their other approach where the auditor does not access the real dataset does not fit our framework.

While synthetic data generation methods often leverage domain knowledge to improve accuracy, we chose to employ a vanilla approach with only the maximum spanning tree method without incorporating additional heuristics. This decision was made due to two main factors: first, the specific domain knowledge about the dataset may not be always readily obvious; second, prediction targets may differ from dataset to dataset, which means that tailored heuristics could not be effectively generalized. By avoiding dataset-specific adjustments, our method remains general and applicable to a wide range of datasets, in addition to being more computationally efficient.


For the measure step, we followed the examples provided in the tool's repository. Gaussian noises are added to the selected marginals. Half of the privacy budget $\epsilon = 1$ is spent on all 1-way marginals and the other half on the selected cliques. These marginals are then fed to the tool to fit the Markov random field. By \cite{mckenna2021winning}, this procedure satisfies $(\alpha,\frac{\alpha}{2 \sigma^2})$-RDP for all $\alpha \geq 1$.

In the COMPAS example, we may select the marginal (sex,race). This marginal is then perturbed by Gaussian noise, as shown in Table~\ref{tab:marginal-perturbed}.

\begin{table}[h]
\caption{Example marginals.}
\label{tab:marginal-perturbed}
\centering
\begin{subtable}[t]{0.45\linewidth}
\centering
\caption{Original marginal.}
\begin{tabular}{lr}
\toprule
Attributes & Count \\
\midrule
Female,African-American & 1537 \\
Female,Asian & 8 \\
Female,Caucasian & 1465 \\
Female,Hispanic & 215 \\
Female,Native American & 15 \\
Female,Other & 143 \\
Male,African-American & 8254 \\
Male,Asian & 63 \\
Male,Caucasian & 4621 \\
Male,Hispanic & 1236 \\
Male,Native American & 42 \\
Male,Other & 717 \\
\bottomrule
\end{tabular}
\end{subtable}
\begin{subtable}[t]{0.45\linewidth}
\centering
\caption{Perturbed marginal.}
\begin{tabular}{lr}
\toprule
Attributes & Count \\
\midrule
Female,African-American & 1545.7885 \\
Female,Asian & 4.3236 \\
Female,Caucasian & 1444.7303 \\
Female,Hispanic & 208.2250 \\
Female,Native American & 26.7479 \\
Female,Other & 150.8175 \\
Male,African-American & 8296.6274 \\
Male,Asian & 92.1107 \\
Male,Caucasian & 4643.0108 \\
Male,Hispanic & 1274.0831 \\
Male,Native American & 28.9542 \\
Male,Other & 710.3473 \\
\bottomrule
\end{tabular}
\end{subtable}
\end{table}

\subsection{Fairness Checking}

After synthesizing the dataset, we used the fairness checker from \cite{yuan2024ensuring} to compute the fairness measures of any incoming AI model.

The fairness checker is an open-sourced public domain Python package that computes various fairness measures, such as those mentioned in Table~\ref{tab:measures}.

The checker is designed to be user-friendly and agnostic to the underlying AI model. It is also designed to be easily extensible to accommodate new fairness measures.

The checker simply iterates through the given database $\db$ and computes the results based on some given predicates on the rows $r_i$s, and finally outputs the fairness measure values.

Protected groups $S$, predicted outcomes $\hat{Y}$, and ground truths $Y$ are all formulated as these predicates. These are straightforward logical boolean expressions. Specifically, they are given as Python functions that output boolean values.

The interpretation of the resulting fairness measure values is dependent on the third-party auditors. The auditors may have different thresholds $\epsilon$ for different fairness measures or different AI models.

In the COMPAS example, the sensitive attribute is sex and the protected group is female, the protected group predicate would be $S := r(sex) = Female$; prediction is if a person will re-offend $\hat{Y} := \mathcal{M}_{recid}(r)$ where $\mathcal{M}$ is the model's prediction of row $r$; ground truth is whether a person does re-offend $Y := r(recid)$. These can be easily implemented in Python as a comparison function.

\section{Experiments}

To test the viability of our method, we compare the metrics computed from the synthetic dataset against those of the original dataset. We used various datasets with fairness concerns mentioned in \cite{pessach2022review}.

We looked at several publicly available datasets, such as Adult\cite{adult_2,Kaggle_Adult_Census_Income}, COMPAS\cite{larson2016propublica,Kaggle_COMPAS_Dataset}, and Diabetes\cite{diabetes_34,Kaggle_Diabetes_Prediction}. These datasets are known to have fairness issues\cite{pessach2022review}.

The fairness checker evaluates datasets based on multiple fairness metrics, such as demographic parity and equalized odds. By comparing these measures between the synthetic and original datasets, we aim to ensure that the synthetic data preserves the fairness properties of the original data. The comparison process is three-fold. It goes as follows.

The dataset is first processed so it can be fed into the synthetic data generator. Some marginals are selected according to their mutual information, as described in the Section~\ref{sec:method}, and the synthetic data generator model is fitted to the original data according to the selected marginals. Then the generator is run 100 times to obtain 100 sets of synthetic data.

Next, several AI models are extracted from various authors from Kaggle. They are finetuned to perform well on the original dataset. For the Adult dataset, we selected three models to experiment: a random forest model is finetuned by hyperparameter search\cite{Ipbyrne2023}; a logistic regression model is finetuned by performing principal component analysis\cite{Prashant1112023}; and a KNN model is manually finetuned\cite{Kothari2020}. For the COMPAS dataset, we selected a random forest model finetuned by hyperparameter search\cite{Sky2024}. For the Diabetes dataset, we selected a random forest model finetuned by over-sampling and under-sampling techniques\cite{Jawat2024}.

The AI models and both the original dataset and synthetic datasets are then fed to the fairness checker. The synthetic datasets were generated 100 times for each model and the fairness scores averaged across the 100 runs. Sensitive attributes are identified based on manual examination with common sense or by referring to \cite{pessach2022review}. Then, applicable fairness measures are computed using the checker for both the original and the synthetic.

This research is conducted in Python Jupyter notebooks and is publicly available. The experiments were conducted on a macOS laptop with an Intel Core i9-9880H CPU (2.3GHz) and 32GB of RAM.

\subsection{Adult Income Dataset}

The Adult\cite{adult_2,Kaggle_Adult_Census_Income} dataset comes from the 1994 census in the United States and contains about 30000 individuals. The dataset has 14 attributes, such as sex, age, and education. The task is to predict whether an individual earns more than \$50,000 a year. We set the sensitive attribute as sex and the protected group as female. We tested three models: a random forest model, a logistic regression model, and a KNN model. The results are shown in Table~\ref{tab:adult_score_rf}, \ref{tab:adult_score_lr}, and \ref{tab:adult_score_knn}. The synthetic data generator fit time is 17 minutes and 1 second.



\begin{table}[h]
\caption{
    Fairness measures experiment results of the Adult dataset on a random forest model.
    Average difference is $0.0775$.
}
\label{tab:adult_score_rf}
\begin{tabular}{lccc}
\toprule
\textbf{Measure} & \textbf{Original} & \textbf{Synthetic} & \textbf{Difference} \\
\midrule
Demographic Parity  & 0.1933 & 0.0114 & 0.1818 \\
Overall Accuracy Equality   & 0.0250 & 0.0169 & 0.0080 \\
Equalized Odds (False Positive)    & 0.0175 & 0.0060 & 0.0115 \\
Equalized Odds (True Positive)    & 0.0114 & 0.0303 & 0.0189 \\
Conditional Use Accuracy Equality (True Positive) & 0.1941 & 0.0315 & 0.1626 \\
Conditional Use Accuracy Equality (True Negative) & 0.1102 & 0.0299 & 0.0803 \\
\bottomrule
\end{tabular}
\end{table}

\begin{table}[h]
\caption{
    Fairness measures experiment results of the Adult dataset on a logistic regression model.
    Average difference is $0.0829$.
}
\label{tab:adult_score_lr}
\begin{tabular}{lccc}
\toprule
\textbf{Measure} & \textbf{Original} & \textbf{Synthetic} & \textbf{Difference} \\
\midrule
Demographic Parity  & 0.0139 & 0.0047 & 0.0092 \\
Overall Accuracy Equality   & 0.0992 & 0.0149 & 0.0843 \\
Equalized Odds (False Positive)    & 0.0017 & 0.0066 & 0.0049 \\
Equalized Odds (True Positive)    & 0.0080 & 0.0072 & 0.0008 \\
Conditional Use Accuracy Equality (True Positive) & 0.3459 & 0.0305 & 0.3154 \\
Conditional Use Accuracy Equality (True Negative) & 0.1120 & 0.0291 & 0.0829 \\
\bottomrule
\end{tabular}
\end{table}

\begin{table}[h]
\caption{
    Fairness measures experiment results of the Adult dataset on a KNN model.
    Average difference is $0.0945$.
}
\label{tab:adult_score_knn}
\begin{tabular}{lccc}
\toprule
\textbf{Measure} & \textbf{Original} & \textbf{Synthetic} & \textbf{Difference} \\
\midrule
Demographic Parity & 0.1721 & 0.0046 & 0.1675 \\
Overall Accuracy Equality &  0.0473 & 0.0128 & 0.0344 \\
Equalized Odds (False Positive) &   0.0571 & 0.0072 & 0.0499 \\
Equalized Odds (True Positive) &   0.1667 & 0.0110 & 0.1557 \\
Conditional Use Accuracy Equality (True Positive) & 0.1009 & 0.0288 & 0.0720 \\
Conditional Use Accuracy Equality (True Negative) & 0.1193 & 0.0319 & 0.0874 \\
\bottomrule
\end{tabular}
\end{table}

\subsection{COMPAS Dataset}

The COMPAS\cite{larson2016propublica,Kaggle_COMPAS_Dataset} dataset comes from an investigative report by ProPublica of the COMPAS criminal recidivism assessment system and contains about 7000 individuals. The dataset has 10 applicable attributes, such sex, age, and priors count. The task is to predict whether an individual will re-offend. We set the sensitive attribute as sex and the protected group as female. The results are shown in Table~\ref{tab:compas_score}. The synthetic data generator fit time is 33 seconds.



\begin{table}[h]
\caption{
    Fairness measures experiment results of the COMPAS dataset on a random forest model.
    Average difference is $0.0727$.
}
\label{tab:compas_score}
\begin{tabular}{lccc}
\toprule
\textbf{Measure} & \textbf{Original} & \textbf{Synthetic} & \textbf{Difference} \\
\midrule
Demographic Parity  & 0.1310 & 0.0809 & 0.0501 \\
Overall Accuracy Equality   & 0.0079 & 0.0146 & 0.0067 \\
Equalized Odds (False Positive)    & 0.0249 & 0.0802 & 0.0553 \\
Equalized Odds (True Positive)    & 0.0177 & 0.0825 & 0.0648 \\
Conditional Use Accuracy Equality (True Positive) & 0.1709 & 0.0551 & 0.1157 \\
Conditional Use Accuracy Equality (True Negative) & 0.1695 & 0.0256 & 0.1439 \\
\bottomrule
\end{tabular}
\end{table}

\subsection{Diabetes Dataset}

The Diabetes\cite{diabetes_34,Kaggle_Diabetes_Prediction} dataset comes from the hospital readmission data published in the 1994 AI in Medicine journal and contains about 100000 individuals. The dataset has 8 attributes, such as gender, age, and smoking history. The task is to predict whether an individual will be readmitted. We set the sensitive attribute as gender and the protected group as female. The results are shown in Table~\ref{tab:diabetes_score}. The synthetic data generator fit time is 1 minute and 24 seconds.




\begin{table}[h]
\caption{
    Fairness measures experiment results of the Diabetes dataset on a random forest model.
    Average difference is $0.0078$.
}
\label{tab:diabetes_score}
\begin{tabular}{lccc}
\toprule
\textbf{Measure} & \textbf{Original} & \textbf{Synthetic} & \textbf{Difference} \\
\midrule
Demographic Parity  & 0.0135 & 0.0015 & 0.0119 \\
Overall Accuracy Equality   & 0.0077 & 0.0011 & 0.0065 \\
Equalized Odds (False Positive)    & 0.0000 & 0.0010 & 0.0010 \\
Equalized Odds (True Positive)    & 0.0086 & 0.0094 & 0.0008 \\
Conditional Use Accuracy Equality (True Positive) & 0.0133 & 0.0059 & 0.0074 \\
Conditional Use Accuracy Equality (True Negative) & 0.0216 & 0.0019 & 0.0197 \\
\bottomrule
\end{tabular}
\end{table}

\section{Evaluation}

Before we delve into the results, we must first understand the nature of the fairness measures. The fairness measures are defined assuming all the information is readily available. This, however, might cause trouble for synthetic data, as they can only work with the information at hand.

By close examination of our scenario, we can observe the following: of the three variables, $(\hat{Y},Y,S)$, two of them, namely $(Y,S)$, are both readily available in the original data. In contrast, $\hat{Y}$ is only available in the output of the given AI model, which cannot be captured by the synthetic data at the time of generation.

Let us inspect the definitions of the fairness measures one by one by expanding them by the definition of conditional probability and marking them according to whether they are in the original data or not; we will box the variables that are not in the original data.

\definecolor{lightgreen}{rgb}{0.6, 1.0, 0.6}
\definecolor{lightred}{rgb}{1.0, 0.6, 0.6}
\newcommand{\mathhighlight}[2]{\colorbox{#1}{$\displaystyle #2$}}
\newcommand{\y}[1]{{#1}}
\newcommand{\n}[1]{\boxed{#1}}

For Demographic Parity, we have
\[
P[\hat{Y} = 1 | S = 1] - P[\hat{Y} = 1 | S \neq 1]
= \frac{P[\n{\hat{Y}} = 1 \cap \y{S} = 1]}{P[\y{S} = 1]} - \frac{P[\n{\hat{Y}} = 1 \cap \y{S} \neq 1]}{P[\y{S} \neq 1]}
\]

For Equalized Odds (False Positive), we have
\[
P[\hat{Y} = 1 | S = 1, Y = 0] - P[\hat{Y} = 1 | S \neq 1, Y = 0]
= \frac{P[\n{\hat{Y}} = 1 \cap \y{S} = 1 \cap \y{Y} = 0]}{P[\y{S} = 1 \cap \y{Y} = 0]} - \frac{P[\n{\hat{Y}} = 1 \cap \y{S} \neq 1 \cap \y{Y} = 0]}{P[\y{S} \neq 1 \cap \y{Y} = 0]}
\]
Equalized Odds (True Positive) follows the same form as Equalized Odds (False Positive).

For Conditional Use Accuracy Equality (True Positive), we have
\[
P[Y = 1 | S = 1, \hat{Y} = 1] - P[Y = 1 | S \neq 1, \hat{Y} = 1]
= \frac{P[\y{Y} = 1 \cap \y{S} = 1 \cap \n{\hat{Y}} = 1]}{P[\y{S} = 1 \cap \n{\hat{Y}} = 1]} - \frac{P[\y{Y} = 1 \cap \y{{S}} \neq 1 \cap \n{\hat{Y}} = 1]}{P[\y{{S}} \neq 1 \cap \n{\hat{Y}} = 1]}
\]
Conditional Use Accuracy Equality (True Negative) follows the same form as Conditional Use Accuracy Equality (True Positive).

For Overall Accuracy Equality, we have
\[
P[Y = \hat{Y} | {S} = 1 ] - P[Y = \hat{Y} | {S} \neq 1]
= \frac{P[\y{Y} = \n{\hat{Y}} \cap \y{S} = 1]}{P[\y{S} = 1]} - \frac{P[\y{Y} = \n{\hat{Y}} \cap \y{{S}} \neq 1]}{P[\y{{S}} \neq 1]}
\]

Here we observed that, of the measures considered, only Conditional Use Accuracy Equality has boxed variables that are in both the numerator and the denominator. This is in contrast to the other measures, which only have one boxed variable in the numerator.

We can say less about fairness with the measures with less information, and thus the measures approximated by synthetic data may be less accurate for Conditional Use Accuracy Equality.

\subsection{Positive Results}

Although the scores may not be fully accurate, and some measures, due to the nature of their definitions, may be harder to approximate, the results showed that synthetic data generally preserves the fairness properties of the original data with a reasonable degree of accuracy.

The average of the differences between the fairness measure values of the original and synthetic datasets is below $0.1$ for all datasets across all models. Different models applied to the same datasets can yield different results, and judging from different metrics, datasets, and models, this indicates that synthetic data is a good approximation of the original data in terms of fairness.

In addition, further experiments showed that, by common sense or in an alternative assumption where the auditor has knowledge of $(S,Y)$ and can add the edge $(S,Y)$ to the Markov random field before fitting, the synthesized data would have been more accurate, even lowering the error of Conditional Use Accuracy Equality in the COMPAS example.

This demonstrated that synthetic data can, to an extent, accurately reflect the fairness properties of the original data. Therefore, fairness analyses performed on synthetic data will yield results mostly consistent with those on the original data.

Auditors hence gained an approach to auditing AI fairness alternative to the traditional method of using real data. When using synthetic data derived from real data, the fairness properties remain meaningful and can still be enough for auditing purposes.

\subsection{Negative Results}

Looking at the actual experimental results, first and foremost, across all datasets, we observed that Conditional Use Accuracy Equality did indeed show higher errors than other measures.

In the Adult random forest experiment, Conditional Use Accuracy Equality (True Positive) has an error of around 0.16. In the COMPAS random forest experiment, both Conditional Use Accuracy Equality measures had an error over 0.1. In the Diabetes random forest experiment, Conditional Use Accuracy Equality (True Negative) has an error much higher than other measures.

This corroborates our assumption that it is harder to approximate sufficiency measures on synthetic data. As there is less information available for the synthesis process, the measures that require more information will be harder to approximate.


This is a fundamental limitation of this approach, and it is important to be aware of this when interpreting the results of fairness analyses. The auditor needs to keep in mind that they are approximating scores, and the scores may not be fully accurate.

\section{Conclusion}

Auditing AI systems with real data can be dangerous due to the potential risks to privacy and the possibility of exposing sensitive information. When evaluating an AI system with personal data, there is a real danger of inadvertently leaking private details about individuals involved in the dataset.

To mitigate this risk, we propose a framework that leverages differentially private synthetic data for fairness audits. Instead of using real data, it generates synthetic data that preserves key patterns of real data but, at the same time, protects individuals' privacy.

The proposed framework, backed by strong privacy guarantees, offers a secure alternative to directly using real data. By applying differential privacy methods, the synthetic data shields individual information from exposure while still enabling thorough fairness evaluations.

Experiments conducted on multiple real-world datasets and across different models show that synthetic data derived from real data retains essential fairness properties. This ensures that any biases present in the original data are captured and reflected in the synthetic counterpart.

While synthetic data is not a flawless substitute for real-world data, it serves as a practical and pragmatic tool for auditing AI fairness in a privacy-preserving manner. Our framework offers an effective means of testing AI systems for fairness while mitigating the privacy concerns surrounding using real data in sensitive applications. Auditors can perform thorough fairness evaluations by generating and using synthetic data while under strong privacy protections.

Future work could focus on applying this framework to larger, more complex datasets, which would help validate the framework's scalability. Furthermore, other privacy-preserving techniques could be explored to complement synthetic data and expand the scope of AI fairness auditing while maintaining privacy. Other means of auditing for certain fairness metrics, such as Conditional Use Accuracy Equality, that are not well suited for synthetic data can be explored. Finally, translating these technical advancements into real-world applications requires policy, legal, and industry collaborations. Research on the legal, ethical, and societal implications of differential privacy and synthetic data will be crucial for broad adoption.


\bibliographystyle{ACM-Reference-Format}
\bibliography{references}


\end{document}